\documentclass[useAMS,usenatbib]{mn2e}

\usepackage{graphicx}

\title[Primordial magnetic field and WMAP large-scale anomalies]
      {Can a primordial magnetic f\/ield originate large-scale anomalies in WMAP data?}

\author[A. Bernui and W. S. Hip\'olito-Ricaldi]{A. Bernui$^{1}$\thanks{E-mail:
bernui@das.inpe.br} and W. S. Hip\'olito-Ricaldi$^{2}$\thanks{E-mail: hipolito@cce.ufes.br} \\
$^{1}$Instituto Nacional de Pesquisas Espaciais, Divis\~{a}o de Astrof\'{\i}sica, 
      Av. dos Astronautas 1758, 12227-010 -- S\~ao Jos\'e dos Campos, SP, Brazil \\
$^{2}$Universidade Federal do Esp\'{i}rito Santo, Departamento de F\'{\i}sica, 
      29060-900 -- Vit\'oria, ES, Brazil}
\begin{document}

\date{Accepted xxxx. Received xxxx; in original form xxxx}

\pagerange{\pageref{firstpage}--\pageref{lastpage}} \pubyear{2008}

\maketitle

\label{firstpage}

%%%%%%%%%%%%%%%%%%%%%%%%%%%%%%%%%%%%%%%%%%%%%%%%%%%%%%%%%%%%%%%%%%%%%%%%%%%%%%%
\begin{abstract}
%%%%%%%%%%%%%%%%%%%%%%%%%%%%%%%%%%%%%%%%%%%%%%%%%%%%%%%%%%%%%%%%%%%%%%%%%%%%%%%
Several accurate analyses of the CMB temperature maps from the Wilkinson Microwave 
Anisotropy Probe (WMAP) have revealed a set of anomalous results, at large angular 
scales, that appears inconsistent with the statistical isotropy expected in the 
concordance cosmological model $\Lambda$CDM. 
Because these anomalies seem to indicate a preferred direction in the space, here 
we investigate the signatures that a primordial magnetic f\/ield, possibly present 
in the photon-baryon f\/luid during the decoupling era, could have produced in the 
large-angle modes of the observed CMB temperature f\/luctuations maps. 
To study these imprints we simulate Monte Carlo CMB maps, which are statistical\/ly 
anisotropic due to the correlations between CMB multipoles induced by the magnetic 
f\/ield. 
Our analyses reveal the presence of the North-South angular correlations asymmetry 
phenomenon in these Monte Carlo maps, and we use these information to establish the 
statistical signif\/icance of such phenomenon observed in WMAP maps. 
Moreover, because a magnetic f\/ield produces planarity in the low-order CMB multipoles, 
where the planes are perpendicular to the preferred direction def\/ined by the 
magnetic f\/ield, we investigate the possibility that two CMB anomalous phenomena, 
namely the North-South asymmetry and the quadrupole-octopole planes alignment, could 
have a common origin. 
Our results, for large-angles, show that the correlations between low-order CMB 
multipoles introduced by a suf\/f\/iciently intense magnetic f\/ield, can reproduce 
some of the large-angle anisotropic features mapped in WMAP data. 
We also reconf\/irm, at more than 95\% CL, the existence of a North-South power 
asymmetry in the WMAP f\/ive-year data. 
\end{abstract}


%%%%%%%%%%%%%%%%%%%%%%%%%%%%%%%%%%%%%%%%%%%%%%%%%%%%%%%%%%%%%%%%%%%%%%%%%%%%%%%
\section{Introduction}\label{Introduction}
%%%%%%%%%%%%%%%%%%%%%%%%%%%%%%%%%%%%%%%%%%%%%%%%%%%%%%%%%%%%%%%%%%%%%%%%%%%%%%%

The f\/ive-year data from the Wilkinson Microwave Anisotropy Probe (WMAP)~\citep{b1,b2,b3,b4,b5}
contain the most valuable cosmological information to study the large-scale properties 
of the universe (see~\citealt{b6,b7,b8,b9,b10,b11,b12}) for previous releases of WMAP data). 
One of these features concerns the hypothesis that the set of temperature f\/luctuations of 
the cosmic microwave background radiation (CMB) is a stochastic realization of a random 
f\/ield, meaning that its angular distribution on the celestial sphere is statistical\/ly 
isotropic at all angular scales.
Examination of the three-year WMAP data conf\/irms highly significant departures 
from statistical isotropy at large angular scales~\citep{b22,b21,b19,b14,b15,b16,b17,%
b20,b13,b18}, previously found also in first-year WMAP 
data~\citep{b33,b34,b23,b24,b31,b25,b26,b27,b28,b29,b32,b30}. 
Evidences for such large-angle anisotropy come from the asymmetry of the CMB angular 
correlations between the northern and southern ecliptic hemispheres (hereafter 
NS-asymmetry), with indications of a preferred axis of maximum hemispherical 
asymmetry~\citep{b23,b24,b25,b26,b27,b28,b29,b16,b30,b13,b18}.
Furthermore, other manifestations of large-angle anisotropy include the unlikely 
quadrupole-octopole planes alignment (refering both to the strong planarity of 
these multipoles as well as to the alignment between such 
planes, see, e.g.,~\citealt{b33,b34,b35,b32,b19,b22,b20,b17}).
Actually, several anomalous results were already reported after analyses of the WMAP 
data~\citep{b37,b38,b31,b36,b39,b40,b41,b107,b42}. 
For a dif\/ferent point of view see, e.g.,~\citet{b43,b44,b45,b46,b47,b48,b49,b50}. 

Possible sources of these anomalies include non-CMB contaminants, like residual 
foregrounds~\citep{b51,b52,b53,b54,b55,b56,b57}, incorrectly  subtracted dipole and/or
dynamic quadrupole terms~\citep{b32,b58,b20}, or systematic errors~\citep{b59,b60}. 
For this, particular ef\/forts have been done by the WMAP team in last releases in 
order to improve the data processing by minimizing the ef\/fects of foregrounds 
(mainly coming from dif\/fuse Galactic emission and astrophysical point-sources), 
artifacts (in the mapmaking process, in the instrument characterization, etc.), and
systematic errors~\citep{b11,b1,b2}. 
As a result, the data released by the WMAP team include the Internal Linear Combination 
(hereafter ILC-5yr) full-sky CMB map suitable for large-angle temperature f\/luctuations 
studies~\citep{b10,b1,b3}. 
Here we investigate the ILC-5yr map, and for completeness, we also study the other 
full-sky cleaned CMB maps that were dif\/ferently processed from WMAP f\/ive- and 
three-year data releases in order to account for foregrounds and systematics. 
Thus, we also consider the Kim-Naselsky-Christensen~\citep{b94}, WMAP-3yr ILC~\citep{b10}, 
de Oliveira-Tegmark~\citep{b53}, and Park-Park-Gott~\citep{b17} CMB maps, hereafter 
termed the HILC-5yr, ILC-3yr, OT-3yr, and PPG-3yr, respectively. 

A number of studies has been done looking for physical explanations of the above 
mentioned anomalies, specially searching for a unifying mechanism relating both the 
NS-asymmetry and the CMB lower multipoles alignment (see, e.g.,~\citealt{b19,b61,b62}). 
Additionally, some processes that breaks down statistical isotropy during the 
inf\/lationary epoch have been suggested~\citep{b63,b64,b65,b66}.
Nonetheless, one can also interpret such CMB large-angle anisotropy as being of 
cosmological origin, in this sense globally axisymmetric space-times have been 
proposed to account for the mapped preferred axis in WMAP 
maps~\citep{b67,b70,b68,b69,b71,b72}.
Previous works investigated the ef\/fects of a primordial homogeneous magnetic f\/ield 
on the CMB temperature f\/luctuations on all angular scales (see, e.g.,~\citealt{b90,b105}). 
Here we study such primordial f\/ield as a possible physical mechanism to produce 
two large-angle phenomena, that is, the CMB NS-asymmetry and the lower CMB multipoles 
alignment.

In section~\ref{magnetic fields} we present this primordial magnetic f\/ield scenario, 
and show the ef\/fect of such a f\/ield on the CMB temperature f\/luctuations.
Then, in order to reveal such ef\/fects on simulated CMB maps we develop a 
geometrical-statistical method, which is presented in section~\ref{method}. 
After that we use our anisotropic indicator to perform, in section~\ref{data analyses}, 
the analyses of both sets of data, the Monte Carlo simulated CMB maps as well as the 
WMAP maps. 
At the end, in section~\ref{conclusions}, we discuss our results and formulate our 
conclusions.

%%%%%%%%%%%%%%%%%%%%%%%%%%%%%%%%%%%%%%%%%%%%%%%%%%%%%%%%%%%%%%%%%%%%%%%%%%%%%%%
\section[]{Primordial magnetic f\/ields scenario}\label{magnetic fields}
%%%%%%%%%%%%%%%%%%%%%%%%%%%%%%%%%%%%%%%%%%%%%%%%%%%%%%%%%%%%%%%%%%%%%%%%%%%%%%%

There is strong observational evidence for the presence of large-scale intergalactic 
magnetic f\/ields of fews $\mu G$, and magnetic f\/ields of similar strength within 
clusters of galaxies (see, e.g.,~\citealt{b73,b74,b75,b76,b77}). 
Nowadays it is believed that these magnetic f\/ields are amplif\/ications of small 
primordial magnetic f\/ields of the order of few nanoGauss, that would have occured 
due to dif\/ferent processes, like galactical dynamo (see, e.g.,~\citealt{b78,b79,b80}), 
during anisotropic protogalactic collapses, or due to dif\/ferential rotation in 
galaxies (see, e.g.,~\citealt{b81,b82}). 
In turn, such primordial seeds of magnetic f\/ields would have several origins, for 
instance an electroweak phase transition~\citep{b84,b83,b85} or quark-hadron phase 
transition (see, e.g.,~\citealt{b86,b87}). 
Here, we assume the existence of a primordial homogeneus magnetic f\/ield and investigate 
their ef\/fects on the CMB temperature f\/luctuations at large-angles. 

According to the Einstein equations for linearized metric perturbations, in absence 
of a magnetic f\/ield, the vector metric perturbations go like $a^{-2}$ and the 
velocity induced by these perturbations goes like $a^{-1}$ (where $a$ is the scale 
factor that accounts for the expansion of the universe). 
Therefore, they decay very fast with the expansion of the universe~\citep{b88} and 
the velocities produced by vector metric perturbations do not contribute 
signif\/icatively to the CMB temperature f\/luctuations.

The presence of a homogeneus magnetic f\/ield in the early universe change this 
scenario because such f\/ields modify the behavior of charged particles in the 
primordial plasma via Lorentz forces producing additional velocity gradients in the 
f\/luid. 
In this way, a magnetic f\/ield induces Alfv\'en waves in the primordial plasma that 
propagate at velocity $v_{\mbox{\rm\footnotesize A}}$, changing the speed of sound 
in the photon-baryon f\/luid as 
$c^2_s \longrightarrow c^2_s + v^2_{\mbox{\rm\footnotesize A}} \cos^2\theta$, 
where~\citep{b89}
\begin{eqnarray}
v_{\mbox{\rm\footnotesize A}}^2 = \frac{B^2_0}{4\pi(\rho + p)} \, ,
\end{eqnarray}
with $\rho$ and $p$ being the density and pressure in the radiation dominated era, 
respectively, $B_0$ is the strength of the magnetic f\/ield $\textbf{B}$, $\theta$ is 
the angle between $\textbf{B}$ and the $\vec{k}$-mode of the Fourier expansion of 
Alfv\'en velocity $v_{\mbox{\rm\footnotesize A}}$. 
These Alfv\'en wave modes induce small rotational velocity perturbations which, for 
the scales of our interest, have the form 
$\vec{v} \, \approx \, \vec{v}_0 \, v_{\mbox{\rm\footnotesize A}}\, k\, t\, \cos \theta$, 
where $t$ is the cosmic time, $k \equiv |\vec{k}|$, and $\vec{v}_0$ is the initial 
velocity, which we assume to have a power spectrum of the form~\citep{b90,b91} 
\begin{eqnarray} \label{espectrov}
\langle v_{0i} \, v_{0j} \rangle \propto k^n \, (\delta_{ij} - k_i \, k_j) \, . 
\end{eqnarray}
Vectorial contributions to CMB temperature f\/luctuations are present via Doppler 
and Sachs-Wolfe ef\/fects~\citep{b90} 
\begin{eqnarray}
\frac{\delta T}{T}(\hat{n})^{\mbox{(\footnotesize vec)}} \,
=\, -\vec{V}_T \cdot \hat{n}|^{t_0}_{t_{\rm\small dec}} \,
+\, \int^{t_0}_{t_{\rm\small dec}} \dot{\vec{\nu}} \cdot \hat{n} 
\, dt \, , 
\end{eqnarray}
where $\hat{n}$ indicates any direction in the sky, $\vec{V}_T = \vec{V} + \vec{v}$,
$\vec{\nu}$ is the vectorial metric perturbation, $\vec{V}$ is the velocity produced
by $\vec{\nu}$, and the integration is between the actual time $t_0$ and the decoupling 
time $t_{\rm\small dec}$.
Thus, only the rotational velocity perturbations $\vec{v}$ contributes ef\/fectively 
to temperature f\/luctuations because $\vec{\nu}$ and $\vec{V}$ decay quickly with the 
expansion of the universe~\citep{b88}. 
Therefore, the signatures of a homogeneus primordial magnetic f\/ield $\textbf{B}$ on 
the CMB temperature f\/luctuations are~\citep{b90} 
\begin{eqnarray} \label{Flutuacoes}
\frac{\delta T}{T}^{\textbf{B}}(\hat{n},k) \,\approx\, 
\hat{n} \cdot \vec{v}_{0} \, v_{A} \, k \, t_{\rm\small dec} \, \cos \theta \, . 
\end{eqnarray}
As we can see, the existence of a preferred direction associated to $\textbf{B}$ 
af\/fects the CMB temperature f\/luctuations through the angle $\theta$, which 
implies, a straight dependence on the orientation of the vector ${\textbf{B}}$, 
consequently, a break down of the statistical isotropy in the CMB sky. 

The most suitable quantity to simulate statistical\/ly anisotropic CMB skies is 
the correlation matrix of multipole coef\/f\/icients. 
For this, we expand the sky temperature f\/luctuations in spherical harmonics 
\begin{eqnarray}\label{Temp} 
\frac{\delta T}{T}(\hat{n}) &=& \sum_{\ell,m} a_{\ell m} Y_{\ell m}(\hat{n})  \, , \\  
a_{\ell m} &=& \int \frac{\delta T}{T}(\hat{n}) \, Y^{\ast}_{\ell m}(\hat{n}) \, 
d\Omega \, , 
\end{eqnarray}
and then use eqs.~(\ref{espectrov}),~(\ref{Flutuacoes}) and~(\ref{Temp}) 
to calculate the correlation matrix $\langle a_{\ell m} a^{\ast}_{\ell' m'} \rangle$. 
In this work we assume a scale-invariant power spectrum for the magnetic f\/ield, 
i.e., $n=-5$ in the eq.~(\ref{espectrov}). 

The explicit calculation of the correlation matrix elements, for a Harrison-Zel'dovich 
scale-invariant power spectrum for the magnetic f\/ield, gives~\citep{b90,b91,b104} 
\begin{eqnarray}\label{CorrelMat1} 
\langle a_{\ell m} a^{\ast}_{\ell' m'} \rangle^{\textbf{B}} 
\!=\! \delta_{m m'} [ \delta_{\ell \ell'} C^{\textbf{B}}_{\ell m} 
+ (\delta_{\ell+1,\ell'-1}+\delta_{\ell-1,\ell'+1}) D^{\textbf{B}}_{\ell m} ],\!\!\!
\end{eqnarray}
where 
\begin{eqnarray}\label{CorrelMat2}
C^{\textbf{B}}_{\ell m} \,=\, 27.12 \times 10^{-16} \left( \frac{B_0}{1\,nG} \right)^4 
\times \nonumber \\
\frac{2\ell^4 + 4\ell^3-\ell^2-3\ell+6m^2-2\ell m^2-2 \ell^2 m^2}
{(\ell+2)(\ell+1)\ell(\ell-1)(2\ell-1)(2\ell+3)} \, ,  
\end{eqnarray}
and
\begin{eqnarray}\label{CorrelMat3}
\frac{D^{\textbf{B}}_{\ell m}}{C^{\textbf{B}}_{\ell m}} 
= \frac{9\pi}{32} \frac{\sqrt{(\ell+m+1)(\ell-m+1)(\ell+m)(\ell-m)}}
{2\ell^4+4\ell^3-\ell^2-3\ell+6m^2-2\ell m^2-2 \ell^2 m^2} \times \nonumber \\ 
\frac{\sqrt{(2\ell-1)(2\ell+3)}(\ell-1)(\ell+2)}{(2\ell+1)} \, , \hspace{1cm}
\end{eqnarray}
noticing that $B_0$ is given in nanoGauss ($nG$). 
Thus, the computation of the correlation matrix leads to non-zero elements of 
the type $(\ell, m)=(\ell', m')$ and $(\ell, m)=(\ell' \pm 2, m')$, while all 
the other elements are zero. 
In other words, the multipole matrix correlation is non-diagonal which is an 
inheritance of its angular dependence on $\theta$ and $\vec{v}_0$, shown 
in the eq.~(\ref{Flutuacoes}), and this fact being a consequence of the presence 
of the magnetic f\/ield at early times. 
In this scenario, correlations appear between dif\/ferent scales, like $\ell$ and 
$\ell \pm 2$,  and then the multipole coef\/f\/icients $a_{\ell m}$ at such scales 
are correlated. 
For this reason, one concludes that large-angle anisotropy features should 
appear in the CMB maps produced by this primordial magnetic f\/ield.
If we consider a magnetic f\/ield in a $\Lambda$CDM universe the correlation 
matrix will be
\begin{eqnarray} \label{Matrixtotal}
\langle a_{\ell m} a^{\ast}_{\ell' m'} \rangle 
\,=\, \langle a_{\ell m} a^{\ast}_{\ell' m'} \rangle^{\mbox{\footnotesize $\Lambda$CDM}} 
+ \langle a_{\ell m} a^{\ast}_{\ell' m'} \rangle^{\textbf{B}} \, ,
\end{eqnarray}
where 
\begin{eqnarray}\label{Clsss} 
\langle a_{\ell m} a^{\ast}_{\ell' m'} \rangle^{\mbox{\footnotesize $\Lambda$CDM}} 
\,=\, C^{\mbox{\footnotesize $\Lambda$CDM}}_{\ell} \,\delta_{m m'} \,\delta_{\ell \ell'} \, . 
\end{eqnarray}
We observe that correlations between temperature f\/luctuations do not only 
depend on the angular separation between two points, but also on their 
orientation with respect to the magnetic f\/ield, that is, $m$-dependence. 
This $m$-dependence, caused by the fact that $B_0 \ne 0$, means that now the 
CMB temperature f\/luctuations are actually statistical\/ly anisotropic. 

Our purpose here is just to illustrate the ef\/fect induced by these large-angle 
anisotropies in CMB maps, and compare them with those features found in the CMB maps 
from WMAP. 
To study the relationship between low-$\ell$ (i.e., from $\ell = 2$ to $\ell = 10$) 
anomalies in such anisotropic magnetic f\/ield scenario, we produce, according to 
eqs.~(\ref{CorrelMat1})--(\ref{Clsss}), f\/ive sets (with dif\/ferent strengths $B_0$) 
of Monte Carlo simulated CMB maps. 
For the statistical\/ly isotropic part (that is, the f\/irst term in 
eq.~(\ref{Matrixtotal})) we use the CMBFAST tool to calculate the angular power 
spectrum $C^{\Lambda \mbox{\sc cdm}}_{\ell}$, and after that we use Cholesky 
decomposition of the matrix (\ref{Matrixtotal}) in order to deal with the 
non-diagonality. 
Then we randomically simulate the $a_{\ell m}$ coef\/f\/icient sets and from them 
we generate the CMB temperature f\/luctuations maps. 
The simulations were performed for magnetic f\/ields intensities $B_0 = (20 \pm 10)\, nG$, 
to be in agreement with known limits at cosmological scales for the $n=-5$ 
power spectrum (see, e.g.,~\citealt{b101,b91,b104}, and~\citealt{b106} 
for other power spectrum indices). 

%%%%%%%%%%%%%%%%%%%%%%%%%%%%%%%%%%%%%%%%%%%%%%%%%%%%%%%%%%%%%%%%%%%%%%%%%%%%%%%%%%%%
\section{The 2PACF and the Sigma-Map method}\label{method}
%%%%%%%%%%%%%%%%%%%%%%%%%%%%%%%%%%%%%%%%%%%%%%%%%%%%%%%%%%%%%%%%%%%%%%%%%%%%%%%%%%%%

Our method to investigate the large-scale angular correlations in CMB temperature 
f\/luctuations maps consists in the computation of the 2-point angular correlation 
function (2PACF)~\citep{b92} in a set of spherical caps covering the 
celestial sphere. 

Let $\Omega_{\gamma_0}^J \equiv \Omega(\theta_J,\phi_J;\gamma_0) 
\subset {\cal S}^2$ be a spherical cap region on the celestial sphere, 
of $\gamma_0$ degrees of aperture, with vertex at the $J$-th pixel, 
$J = 1, \ldots, N_{\mbox{\footnotesize caps}}$, where $(\theta_J,\phi_J)$ 
are the angular coordinates of the $J$-th pixel's center.
Both, the number of spherical caps $N_{\mbox{\footnotesize caps}}$ and the 
coordinates of their centers $(\theta_J,\phi_J)$ are def\/ined using the HEALPix 
pixelization scheme~\citep{b93}.  The union of the $N_{\mbox{\footnotesize caps}}$ 
spherical caps covers completely the celestial sphere ${\cal S}^2$. 

Given a pixelized CMB map, the 2PACF of the temperature f\/luctuations 
$\delta T$ corresponding to the pixels located in the spherical cap 
$\Omega_{\gamma_0}^J$ is def\/ined by~\citep{b92} 
\begin{equation} \label{2PACF}
\mbox{\rm C}(\gamma)^J \equiv 
\langle\, \delta T(\theta_i,\phi_i) \delta T(\theta_{i'},\phi_{i'}) \,\rangle \, ,
\end{equation}
where 
$\cos\gamma = \cos\theta_i \cos\theta_{i'} 
+ \sin\theta_i \sin\theta_{i'} \cos(\phi_i\!-\!\phi_{i'})$, 
and $\gamma \in (0,2\gamma_0]$ is the angular distance between the $i$-th and 
the $i'$-th pixels centers. The average $\langle \,\, \rangle$ in the above
equation is done over all the products 
$\delta T(\theta_i,\phi_i) \delta T(\theta_{i'},\phi_{i'})$ such that 
$\gamma_k \equiv \gamma \in ((k-1)\delta,\, k\delta]$, for 
$k = 1,..., N_{\mbox{\footnotesize bins}}$, where 
$\delta \equiv 2\gamma_0 / N_{\mbox{\footnotesize bins}}$
is the bin-width. 
We denote by $\mbox{\rm C}_k^J \equiv \mbox{\rm C}(\gamma_k)^J$ the value of the 
2PACF for the angular distances $\gamma_k \in ((k-1)\delta,\, k\delta]$. 
%
%%%%%%%%%%%%%%%%%%%%%%%%%%%%%%%%  sigma-map  %%%%%%%%%%%%%%%%%%%%%%%%%%%%%%%
%
Def\/ine now the scalar function 
$\sigma: \Omega_{\gamma_0}^J \subset {\cal S}^2 \mapsto {\Re}^{+}$, 
for $J = 1, \ldots, N_{\mbox{\footnotesize caps}}$, which assigns to the 
$J$-cap, centered at $(\theta_J,\phi_J)$, a real positive number 
$\sigma_J \equiv \sigma(\theta_J,\phi_J) \in \Re^+$. 
The most natural way of def\/ining a measure $\sigma$ is through the 
variance of the $\mbox{\rm C}_k^J$ function~\citep{b18},  
\begin{equation} \label{sigma}
\sigma^2_J  \equiv \frac{1}{N_{\mbox{\footnotesize bins}}}
\sum_{k=1}^{N_{\mbox{\footnotesize bins}}} (\mbox{\rm C}_k^J)^2 \,\, . 
\end{equation}
To obtain a quantitative measure of the angular correlations in a CMB map, we 
cover the celestial sphere with $N_{\mbox{\footnotesize caps}}$ spherical caps, and 
calculate the set of sigma values $\{ \sigma_J, \, J=1,...,N_{\mbox{\footnotesize caps}} \}$ 
using the eq.~(\ref{sigma}). 
Associating the $J$-th sigma value $\sigma_J$ to the $J$-th pixel, for 
$J=1, \ldots, N_{\mbox{\footnotesize caps}}$,  one f\/ills the celestial sphere 
with positive real numbers, and according to a linear scale (where 
$\sigma^{\mbox{\footnotesize minimum}} \rightarrow blue$,  
$\sigma^{\mbox{\footnotesize maximum}} \rightarrow red$), one converts this numbered 
map into a coloured map: this is the sigma-map. 
F\/inally, we f\/ind the multipole components of a sigma-map by calculating its 
angular power spectrum. 
In fact, given a sigma-map one can expand $\sigma = \sigma(\theta,\phi)$ in spherical 
harmonics: 
$\sigma(\theta,\phi) = \sum_{\ell,\, m} A_{\ell\, m} Y_{\ell\, m}(\theta,\phi)$. 
Then the set of values $\{ S_{\ell},\, \ell=1,2,... \}$, where 
$S_{\ell} \equiv (1 / (2\ell+1)) \sum_{m={\mbox{\small -}}\ell}^{\ell} \, |A_{\ell\, m}|^2$, 
give the angular power spectrum of the sigma-map.

A power spectrum $S_{\ell}^{\mbox{\footnotesize WMAP}}$ of a sigma-map computed 
from a given WMAP map, provides quantitative information about large-angle anisotropy 
features of such a CMB map when compared with the mean of sigma-map power spectra 
obtained from Monte Carlo CMB maps produced under the statistical isotropy hypothesis. 
As we shall see, the sigma-map analysis is able to reveal large-angle anisotropies 
such as the NS-asymmetry.

%%%%%%%%%%%%%%%%%%%%%%%%%%%%%%%%%%%%%%%%%%%%%%%%%%%%%%%%%%%%%%%%%%%%%%%%%%%%%%%% 
\section{Data analyses and results}\label{data analyses}
%%%%%%%%%%%%%%%%%%%%%%%%%%%%%%%%%%%%%%%%%%%%%%%%%%%%%%%%%%%%%%%%%%%%%%%%%%%%%%%% 

Now we shall apply the sigma-map method to scrutiny the large-angle correlations 
present in the WMAP maps, and perform a quantitative comparison with the result 
of a similar analysis performed in sets of simulated Monte Carlo (MC) CMB maps. 
These sets of simulated maps are produced according to the primordial magnetic 
f\/ield scenario discussed above, where we consider f\/ive cases, for an equal number 
of values for the parameter $B_0$, to examine the inf\/luence of the f\/ield intensity 
on the CMB angular correlations. 
These analyses let us to test the hypothesis that the anomalous angular correlations 
found in the WMAP data could be explained by the presence of a magnetic f\/ield acting 
in the decoupling era. 
The WMAP data under investigation are the full-sky cleaned CMB maps derived from 
WMAP f\/ive- and three-year releases, namely the ILC-5yr, HILC-5yr, ILC-3yr, OT-3yr, 
and PPG-3yr CMB maps. 
Because we are interested in understand the possible cause-ef\/fect relationship 
between a primordial magnetic f\/ield and the large-angle anomalies mapped in CMB 
WMAP data, we concentrate our study on the low-order multipoles range $\ell = 2 - 10$.

For our analyses we produce f\/ive sets of $1\,000$ MC CMB maps each, 
corresponding to the cases when $B_0 = 0,\,10,\,15,\,20$, and $30\, nG$. 
The case $B_0 = 0$ means that the MC maps were generated using a pure 
$\Lambda$CDM angular power spectrum seed~\citep{b12,b5}, and this 
case refers to the statistical\/ly isotropic CMB maps. 

\begin{figure}  
\includegraphics[width=8cm, height=7cm]{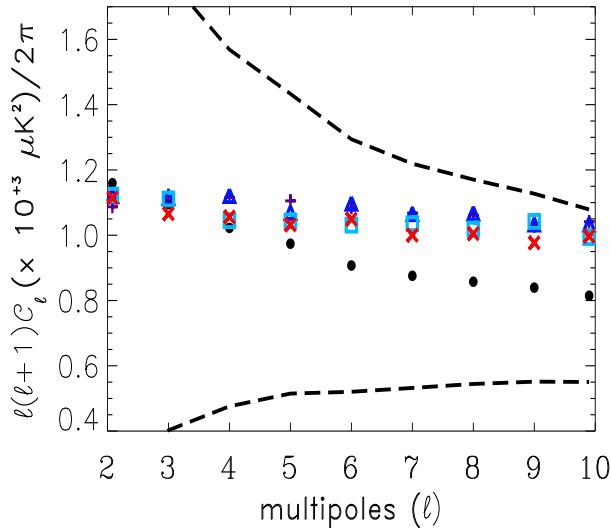}
\caption{\label{fig1}
Normalized mean angular power spectra of the MC CMB maps used to produce 
the sigma-maps-MC. 
The bullet, plus, triangle, square, and {\Large $\times$} symbols represent the data  
when $B_0 = 0,\, 10,\, 15,\, 20,\, 30\, nG$, respectively. 
The dashed lines are the cosmic variance limits of the $\Lambda$CDM 
(i.e., $B_0=0$) power spectrum case. 
Note that the four sets of MC maps with $B_0 \ne 0$ have their angular power 
spectra consistent with the $\Lambda$CDM case.} 
\end{figure}

For $B_0 \ne 0$ the simulated MC CMB maps were produced considering the two 
contributions to the random $a_{\ell m}$-modes according to eq.~(\ref{Matrixtotal}), 
that is, the statistical\/ly isotropic part plus the component due to the magnetic 
f\/ield. 
As a result of this, the mean angular power spectra of each set of MC maps, 
for $B_0 =10,\,15,\,20,\,30\,nG$, satisf\/ies the Sachs-Wolfe plateau ef\/fect. 
However, due to the contributions 
$\langle a_{\ell m} a^{\ast}_{\ell' m'} \rangle^{\textbf{B}}$ (see the second term 
in eq.~(\ref{Matrixtotal})) the plateau of these mean angular power spectra are 
shifted upwards, where the value of such shifts is proportional to $B_0$. 
In order to compare features obtained, through the sigma-map method, from MC maps 
with power spectra that be consistent with the $\Lambda$CDM power spectrum we 
normalize the MC power spectra corresponding to the $B_0 \ne 0$ cases. 
The normalized mean angular power spectra $C_{\ell}$ of these four sets of MC maps 
$B_0 \ne 0$ cases and the angular power spectrum of the $\Lambda$CDM model, plus 
its cosmic variance limits, are shown in f\/igure~\ref{fig1}. 

In our simulations we have assumed that the magnetic f\/ield is pointing in the South 
Galactic Pole--North Galactic Pole (SGP--NGP) direction. Notice that all the 
sky maps plotted here are in galactic coordinates, which means that the equator of 
the map corresponds to the Galactic plane, and the axis SGP--NGP is perpendicular 
to this plane. 

%%%%%%%%%%%%%%%%%%%%%%%%%%%% FIGURA 1 %%%%%%%%%%%%%%%%%%%%%%%%%%%%%%%%%%%%%%%%%%%%
%%%%%%%%%%%%%%%%%%%%%%%%%%%%% FIGURA 1 %%%%%%%%%%%%%%%%%%%%%%%%%%%%%%%%%%%%%%%%%%%%

\begin{figure}  
\hspace{0.9cm}
\includegraphics[width=6.5cm,height=12cm]{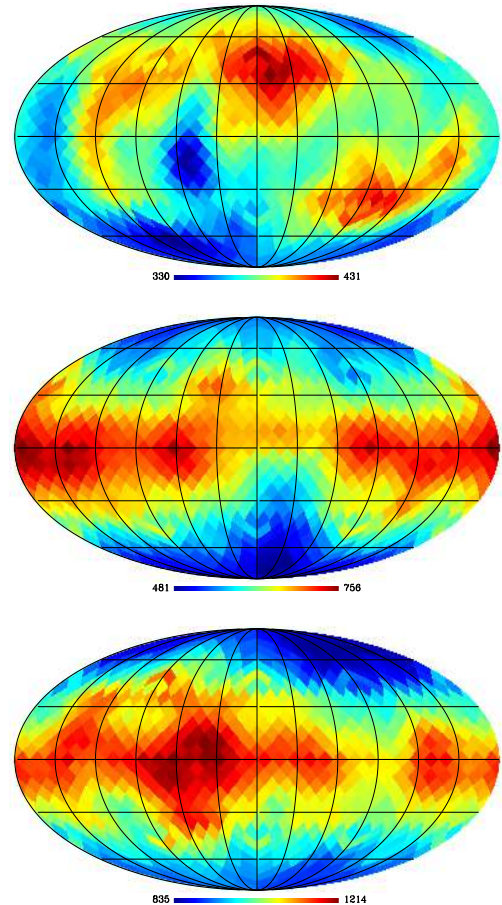}
\caption{\label{fig2}
These sky maps, in Galactic coordinates and from top to bottom, represent the mean 
of 100 sigma-maps-MC obtained from a similar number of MC computed considering 
three cases: $B_0= 0$, $B_0= 10\, nG$, and $B_0= 30\, nG$, respectively, and with the 
magnetic f\/ield pointing in the SGP--NGP direction, which means that the equator 
is the preferred plane. 
Notice that for $B_0 \ne 0$ the region around the equator concentrates strong 
angular correlations, here represented by the intense and large red sky patches.} 
\end{figure}

\begin{figure}
\hspace{0.9cm}
\includegraphics[width=6.5cm,height=12cm]{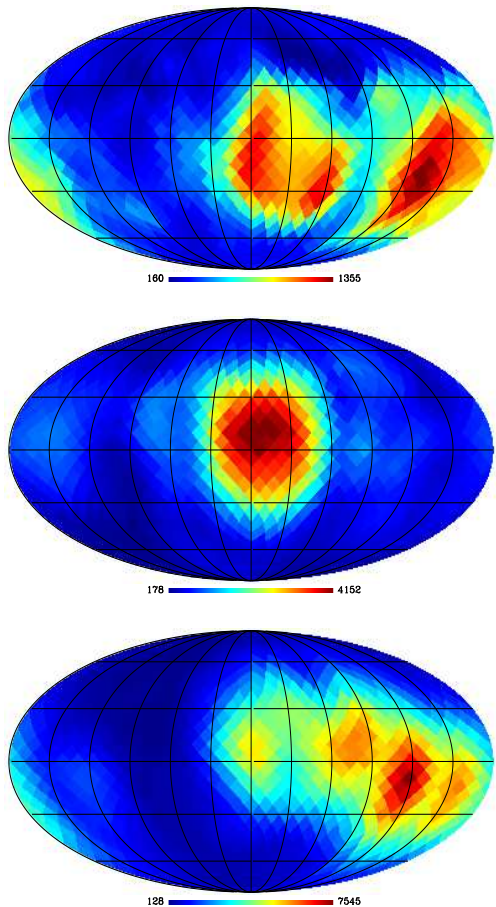}
\caption{\label{fig3} 
For illustration we show three sigma-maps: 
the sigma-map-WMAP from the ILC-5yr CMB map (top), and two sigma-maps-MC having 
large dipole term $S_{1}$ one obtained from a MC map with $B_0=10\, nG$ (middle) 
and the other obtained from a MC map with $B_0=30\, nG$ (bottom).}
\end{figure}

The data analyses consist on the following steps. 
For each MC temperature map we compute its corresponding sigma-map (hereafter 
called sigma-map-MC), and then we calculate its angular power spectrum 
$\{ S_{\ell}, \ell=1,2,... \}$. 
The statistical signif\/icance of the sigma-map angular power spectra computed from WMAP 
data (hereafter called sigma-map-WMAP) comes from the comparison with the f\/ive sets of 
1\,000 angular power spectra obtained from the sigma-maps-MC. 
To illustrate the ef\/fect of the magnetic f\/ield in the CMB low-order multipoles in 
MC maps, we show three cases in f\/igure~\ref{fig2}: 
the mean of 100 sigma-maps-MC obtained from a similar number of MC computed considering 
the statistical\/ly isotropic case $B_0 = 0$ (top), and considering magnetic f\/ield 
intensities $B_0 = 10\, nG$ (middle) and $B_0 = 30\, nG$ (bottom) respectively. 
In f\/igure~\ref{fig3} instead, we show three sigma-maps with large value of the 
dipole term $S_{1}$, indicative of the hemispherical asymmetry phenomenon, produced 
from the ILC-5yr map (top) and from MC CMB maps with dif\/ferent magnetic f\/ield 
intensities: $B_0=10\, nG$ (middle) and $B_0=30\, nG$ (bottom).

%%%%%%%%%%%%%%%%%%%%%%%%%%%%% FIGURA 2 %%%%%%%%%%%%%%%%%%%%%%%%%%%%%%%%%%%%%%%%%%%
%%%%%%%%%%%%%%%%%%%%%%%%%%%%% FIGURA 2 %%%%%%%%%%%%%%%%%%%%%%%%%%%%%%%%%%%%%%%%%%%

%%%%%%%%%%%%%%%%%%%%%%%%%%%%% FIGURA 3 %%%%%%%%%%%%%%%%%%%%%%%%%%%%%%%%%%%%%%%%%%%
%%%%%%%%%%%%%%%%%%%%%%%%%%%%% FIGURA 3 %%%%%%%%%%%%%%%%%%%%%%%%%%%%%%%%%%%%%%%%%%%

\begin{figure}  
\hspace{-0.5cm}
\includegraphics[width=9cm, height=8cm]{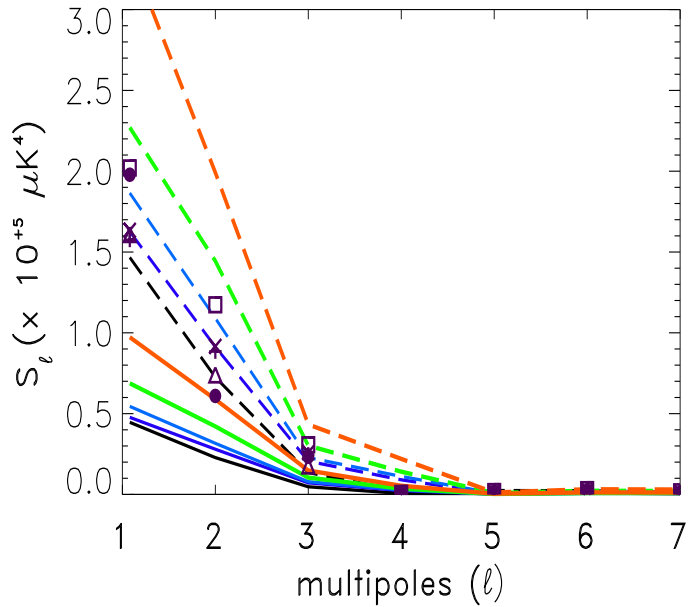}
\caption{\label{fig4}
Angular power spectra of the sigma-maps-MC for MC produced with dif\/ferent 
magnetic f\/ields intensities together with the sigma-maps-WMAP spectra. 
The plus, square, {\Large $\times$}, bullet, and triangle symbols represent the 
data from the ILC-5yr, HILC-5yr, ILC-3yr, PPG-3yr, and OT-3yr CMB maps, 
respectively. 
From bottom to top, the solid (dashed) lines represent the mean (95\% CL) 
values computed from the set of sigma-maps-MC corresponding to the cases 
$B_0 = 0,\, 10,\, 15,\, 20,\, 30\, nG$, respectively.} 
\end{figure}

In f\/igure~\ref{fig4} we present the results of the sigma-maps spectra analyses. 
We observe that the sigma-maps-WMAP, obtained from the ILC-5yr, HILC-5yr, ILC-3yr, 
OT-3yr, and PPG-3yr CMB maps, reveal a dipole moment $S_{1}^{\mbox{\sc wmap}}$ 
larger than 95\% of the values $S_{1}^{\mbox{\sc mc}-\Lambda \mbox{\sc cdm}}$, 
corresponding to sigma-maps-MC obtained from statistical\/ly isotropic CMB maps. 
This fact indicates that the NS-asymmetry phenomenon is present in WMAP data 
at 95\% CL. 
However, we also observe in f\/igure~\ref{fig4} that the angular power spectra of 
the sigma-maps-MC varies with the magnetic f\/ield intensity, thus to larger values 
of $B_0$ correspond sigma-maps-MC with larger values of the terms $S_1$ (dipole), 
$S_2$ (quadrupole), $S_3$ (octopole), etc. 
In other words, a larger value of $B_0$ produces a stronger hemispherical asymmetry 
in the MC CMB maps. 
Therefore, we conclude that for suf\/f\/iciently large $B_0$, one can interpret 
the spectra resulting from sigma-maps-WMAP as being not anomalous at all, but 
consistent with those CMB temperature maps produced according to the primordial 
magnetic f\/ield scenario. 
Interestingly, simulations seems to indicate that an intensity of $B_0 \sim 15\, nG$ 
is enough to have a NS-asymmetry as strong as WMAP data (see f\/igure~\ref{fig4}). 

%%%%%%%%%%%%%%%%%%%%%%%%%%% FIGURA 4 %%%%%%%%%%%%%%%%%%%%%%%%%%%%%%%%%%%%%%%%%%%%
%%%%%%%%%%%%%%%%%%%%%%%%%%% FIGURA 4 %%%%%%%%%%%%%%%%%%%%%%%%%%%%%%%%%%%%%%%%%%%%

The ef\/fect of the magnetic f\/ield on the set of $(\ell,\,m)$-modes, for a given 
multipole $\ell$, deserves a close inspection. 
We know that in the statistical\/ly isotropic case the power of the $\ell$-multipole 
in a CMB map, given by $C_{\ell} = (1 / 2\ell+1) \sum_{m=-\ell}^{\ell} |a_{\ell m}|^2$, 
is uniformly distributed between the $(2\ell + 1)$ modes, for $m = -\ell,...0,...\ell$. 
However, when a primordial magnetic f\/ield is acting on the CMB, and pointing in 
the SGP-NGP direction, we expect a non-uniform distribution of such power with 
a net predilection for the planar modes 
perpendicular to this axis, event that is termed the $m = \ell$-preference. 

To investigate if the multipoles of our MC are experiencing this planarity phenomenon 
when $B_0 \ne 0$, we perform a probabilistic analysis of how the power of the 
quadrupole moment $C_2$ and octopole moment $C_3$ are distributed in their 
$(2\ell + 1) = 5$ and $(2 \ell + 1) = 7$ modes, respectively.  
For this, as a criterium for measuring the predilection for the $(\ell, m)$-mode 
we consider those values obtained for the quadrupole and octopole from the ILC-5yr 
CMB map, that is, $(\ell, m)$ is a preferred mode when it takes more than 35\% 
of the power of such multipole, i.e., 
$|a_{\ell m}|^2 + |a_{\ell m}^{\ast}|^2 > 0.35\, (2\ell + 1)\, C_{\ell}$. 
Given a subset of MC where the power of at least one of their $(2\ell' + 1)$-modes, 
of a given $\ell'$-multipole, is greater than $0.35\, (2\ell' + 1)\, C_{\ell'}$, 
we def\/ine ${\cal P}_{\ell' m'}$ as the probability that the mode $(\ell', m')$ 
satisf\/ies the power distribution criterium (PDC): 
$|a_{\ell' m'}|^2 + |a_{\ell' m'}^{\ast}|^2 > 0.35\, (2\ell' + 1)\, C_{\ell'}$. 
For instance, ${\cal P}_{2,2}$ is the probability that the quadrupolar mode 
$(\ell', m')=(2,2)$ satisf\/ies $|a_{2,2}|^2 + |a_{2,2}^{\ast}|^2 > (0.35)\,(5)\,C_{2}$, 
where such probability is computed considering the subset of MC where at least one 
of the 5 modes $(\ell', m')$, for $\ell'=2$ and $m'=-2,-1,0,1,2$, 
has power greater than $(0.35)\, (5)\, C_{2}$.
We perform a comparative analysis for two sets of MC data, namely the set of 1\,000 MC 
$\Lambda$CDM (i.e., $B_0 = 0$, hereafter MC-$\Lambda$CDM) and the set of 1\,000 MC 
produced with $B_0 = 20\, nG$ (hereafter MC-B20). 

Additionally, to realize a possible correlation between the quadrupole-octopole 
planes alignment and the NS-asymmetry phenomena we also investigate the occurrence 
of the $m = \ell$-preference in three subsets of the MC-$\Lambda$CDM and MC-B20 sets,
namely those subsets that satisfies the PDC for the quadrupole and the octopole 
simultaneously and such that their corresponding sigma-maps-MC have: 
(i) any dipole moment value $S_{1}^{\mbox{\sc mc}}$, hereafter denoted by 
MC-$\Lambda$CDM 0-SD and MC-B20 0-SD subsets, respectively; 
(ii) dipole moment value $S_{1}^{\mbox{\sc mc}}$ larger than the mean value plus 
one standard deviation, hereafter these data are termed MC-$\Lambda$CDM 1-SD and 
MC-B20 1-SD, respectively;
(iii) dipole moment value $S_{1}^{\mbox{\sc mc}}$ larger than the mean value plus 
two standard deviations, hereafter these data are termed MC-$\Lambda$CDM 2-SD and 
MC-B20 2-SD, respectively. 
The fraction of MC, in the MC-$\Lambda$CDM and MC-B20 sets, that satisf\/ies the above 
PDC for the quadrupole and the octopole simultaneously is $\sim 0.63$ for all these 
data subsets. 

\begin{figure}
\includegraphics[width=8.5cm,height=10cm]{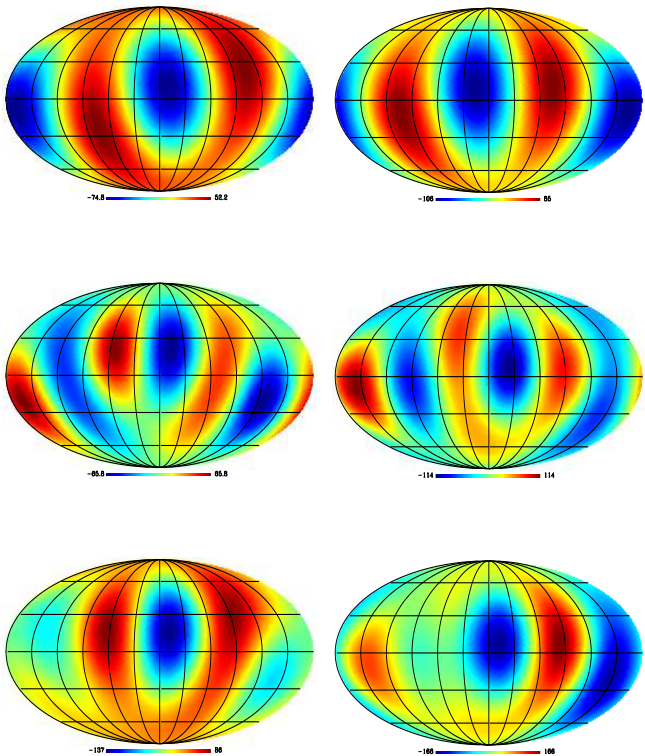}
\caption{\label{fig5} 
The left (right) panels are the quadrupole $\ell=2$, the octopole $\ell=3$ 
and their sum $\ell=2+3$ corresponding to the MC CMB map that produces the 
sigma-map-MC plotted in the middle (bottom) panel of f\/igure~\ref{fig3}.} 
\end{figure}

Our results for these computations are shown in Table~\ref{table1}, where we 
observe the following results. 
F\/irst, as expected for the three sets of MC-$\Lambda$CDM data, the analysis 
reveals that the quadrupole and the octopole have their power uniformly distributed 
between all the $(\ell, m)$-modes. 
Second, regarding the MC-B20 data sets, it is observed a weak preference for the 
$(\ell, m) \ne (\ell, 0)$ modes in 0-SD and 1-SD data sets, while the planarity or 
$m \!=\! \ell$-preference is actually evident in the set 2-SD. 
Due to this fact, and according to the def\/inition of ${\cal P}_{\ell m}$, we conclude 
that there is a net correlation between the quadrupole-octopole planes alignment 
(represented by the $m=\ell$-preference through the large values of ${\cal P}_{2, 2}$ 
and ${\cal P}_{3, 3}$) and the NS-asymmetry phenomena (represented by the fact that 
these large probability values appear only considering those MC that produce 
sigma-maps-MC with the largest dipoles $S_{1}^{\mbox{\sc mc}}$). 
We understand this $m = \ell$-preference as a consequence of the planarity induced 
by the preferred direction settled by the magnetic f\/ield, as illustrated with two 
examples in f\/igure~\ref{fig5}. 
Additionally, one notices the ef\/fect that this planarity produces in the sigma-maps-MC 
causing a concentration of strong angular correlations (red regions) around the equator, 
which is the preferred plane, as clearly seen in the mean of the sigma-maps-MC skies 
showed in f\/igure~\ref{fig2}. 

%%%%%%%%%%%%%%%%%%%%%%%%%%%%%%%%%  TABELA 1  %%%%%%%%%%%%%%%%%%%%%%%%%%%%%%%%%%%%%
\begin{table*}
\caption{\label{table1} 
Probability comparative analyses of the power distribution for the $a_{\ell m}$-modes 
of the quadrupole ($\ell=2$) and the octopole ($\ell=3$) in those subsets of 
MC-$\Lambda$CDM and MC-B20 that satisfies the PDC for the quadrupole and octopole 
simultaneously. 
The probability ${\cal P}_{\ell m}$ is def\/ined in the text. 
0-SD means that we are consider for analysis the subset of the MC sets (mentioned above) 
such that their corresponding sigma-maps-MC have any dipole moment value 
$S_{1}^{\mbox{\sc mc}}$; 
1-SD (2-SD) means that we are consider for analysis the subset of the MC sets (mentioned 
above) such that their corresponding sigma-maps-MC have dipole moment value 
$S_{1}^{\mbox{\sc mc}}$ larger than the mean value plus one (two) standard deviation(s).} 
\begin{tabular}{cccccccc} 
\hline 
MC data $\backslash$ ${\cal P}_{\ell m}$ & ${\cal P}_{2,0}$ & ${\cal P}_{2,1}$ & ${\cal P}_{2,2}$ 
& ${\cal P}_{3,0}$   & ${\cal P}_{3,1}$ & ${\cal P}_{3,2}$ & ${\cal P}_{3,3}$        \\ \hline
MC-$\Lambda$CDM 0-SD & 26.7\% & 39.0\% & 34.3\% & 26.3\% &  23.8\% & 26.6\% & 23.3\% \\
MC-$\Lambda$CDM 1-SD & 25.8\% & 39.9\% & 34.3\% & 25.9\% &  22.4\% & 27.9\% & 23.8\% \\
MC-$\Lambda$CDM 2-SD & 30.2\% & 32.6\% & 37.2\% & 20.2\% &  26.6\% & 26.6\% & 26.6\% \\
MC-B20 0-SD          & 10.4\% & 45.0\% & 44.6\% & 13.0\% &  29.7\% & 27.6\% & 29.7\% \\
MC-B20 1-SD          & 11.4\% & 39.3\% & 49.3\% &  7.5\% &  31.9\% & 25.3\% & 35.3\% \\
MC-B20 2-SD          &  4.0\% & 32.0\% & 64.0\% &  0.0\% &  14.3\% & 28.6\% & 57.1\% \\
\hline
\end{tabular}
\end{table*}
%%%%%%%%%%%%%%%%%%%%%%%%%%%%%%%%%  TABELA 1  %%%%%%%%%%%%%%%%%%%%%%%%%%%%%%%%%%%%%

%%%%%%%%%%%%%%%%%%%%%%%%%%%%%% FIGURA 5 %%%%%%%%%%%%%%%%%%%%%%%%%%%%%%%%%%%%%%%%%%
%%%%%%%%%%%%%%%%%%%%%%%%%%%%%% FIGURA 5 %%%%%%%%%%%%%%%%%%%%%%%%%%%%%%%%%%%%%%%%%%

A important part of our analyses are the robustness tests. 
To assert the robustness of our results when using a dif\/ferent set of parameters 
in the sigma-map method, we investigated the ef\/fect of changing 
$\gamma_0, \, N_{\mbox{\footnotesize bins}}, \, N_{\mbox{\footnotesize caps}}$, 
and $N_{\mbox{\footnotesize side}}$ of the CMB map in analysis, in 
the computation of the sigma-maps. 
For this we performed several sigma-map calculations using spherical caps of 
$\gamma_0 = 45^{\circ},\, 60^{\circ}$ of aperture, 
$N_{\mbox{\footnotesize bins}}=45,\, 60,\, 90$, and 
$N_{\mbox{\footnotesize caps}}=768,\, 3\,072$, resulting in minor dif\/ferences 
in all these cases. 
Additionally, we also examine the inf\/luence of the angular resolution of the 
CMB maps by computing the sigma-maps considering dif\/ferent pixelisation 
parameters of the CMB maps, namely $N_{\mbox{\footnotesize side}}=16$ and $32$. 
In particular, the sigma-maps showed in f\/igure~\ref{fig2}, plotted in galactic 
coordinates, and their corresponding angular power spectra analyses in f\/igure~\ref{fig1}, 
were calculated using $\gamma_0 = 45^{\circ}$, 
$N_{\mbox{\footnotesize bins}}=45$, $N_{\mbox{\footnotesize caps}}=768$, and 
$N_{\mbox{\footnotesize side}}=32$. 
Summarizing, our robustness tests show that using a set of parameters within a 
certain range of values, we obtain results that are fully consistent with those 
showed in f\/igure~\ref{fig4}. 

%%%%%%%%%%%%%%%%%%%%%%%%%%%%%%%%%%%%%%%%%%%%%%%%%%%%%%%%%%%%%%%%%%%%%%%%%%%%%%%%
\section{Conclusions}\label{conclusions}
%%%%%%%%%%%%%%%%%%%%%%%%%%%%%%%%%%%%%%%%%%%%%%%%%%%%%%%%%%%%%%%%%%%%%%%%%%%%%%%%

We investigated a plausible primordial scenario where a homogeneous magnetic 
f\/ield acting on the photon-baryon f\/luid at the recombination era, introduces 
a preferred direction that establish a statistical isotropy breaking. 
This setting of\/fers a possible explanation for those anomalies found in WMAP data 
that seems to be associated to a preferred axis in the space, like the hemispherical
NS-asymmetry  and the planarity of some CMB low-order multipoles (where such
a plane is  perpendicular to that axis). It is found that this magnetic f\/ield
induces correlations between the CMB  multipoles, manifested through non-diagonal
terms in the multipole correlation matrix (see 
eqs.~(\ref{CorrelMat1})--(\ref{CorrelMat3})). 
Accordingly, we have simulated f\/ive sets of CMB skies considering an equal number
of  strengths $B_0$ of the magnetic f\/ield. 
These MC CMB maps, with multipoles in the range $\ell = 2 - 10$, are used to 
investigate their large-angle correlations using our sigma-map method. 
With these data sets we performed a quantitative analysis of their large-angle 
signatures and perform a comparison with similar features computed from WMAP maps. 
For this, we consider a set of full-sky cleaned CMB maps, obtained from f\/ive-year 
and three-year WMAP data releases, namely the ILC-5yr, KCN-5yr, ILC-3yr, OT-3yr, 
and PPG-3yr maps, which were dif\/ferently processed in order to account for 
foregrounds and systematic errors. 

Our results can be summarized as follows. 
F\/irst, our analyses corroborate that an uneven hemispherical distribution in the 
power of the large-angular correlations, best known as NS-asymmetry phenomenon, 
is present in all these WMAP maps at more than 95\% CL as compared with statistical\/ly 
isotropic CMB maps produced according to the $\Lambda$CDM cosmological model (i.e., 
$B_0=0$). 
Second, we found that this hemispherical asymmetry phenomenon, is such that higher is 
the value of the f\/ield strenght $B_0$, greater is the probability that such asymmetry 
be present in the MC CMB maps, and be revealed with a high dipole value 
$S_{1}^{\mbox{\sc mc}}$ in its corresponding sigma-map-MC. 
In other words, our results show that the correlations introduced by a magnetic f\/ield 
mechanism in low-order CMB multipoles, for a suf\/f\/iciently intense f\/ield, reproduce 
the NS-asymmetry present in WMAP data as a common phenomena and not as an anomalous one. 
Third, as expected, the MC CMB maps with $B_0 \ne 0$ exhibit the planarity ef\/fect 
in their low-order multipoles (see, for illustration, f\/igure~\ref{fig5}) as 
quantif\/ied in the Table~\ref{table1}, where the $\ell$-multipole power is concentrated 
in the $m = \pm \ell$ modes, and as evidenced by the intense red spots representative 
of strong angular correlations appearing in the equatorial region of the sigma-maps-MC 
(see, for il\/lustration, f\/igure~\ref{fig2}). 
Fourth, as shown in the Table~\ref{table1}, we found a significant correlation 
between the NS-asymmetry and quadrupole-octopole (planarity and) alignment phenomena. 
For completeness, we have also verif\/ied that our results are robust under dif\/ferent 
sets of parameters, as mentioned in the previous section, involved in the sigma-map 
method calculations. 

The large angular scales CMB anomalies challenge the statistical isotropy expected 
in the $\Lambda$CDM concordance model. 
Our results suggests that perhaps some of them could be suitable explained by 
just one physical phenomenon. 
As a matter of fact, there are several attempts to f\/ind the origin of large-angle 
CMB anomalies, but further investigations are needed to fully comprehend if they 
have one or more causes. 
In this sense, statistical\/ly anisotropic models explaining several anomalies with 
a minimum set of hipothesis should be explored. 

We believe that our results shall motivate the study of other statistical\/ly 
anisotropic scenarios, in particular, those where correlations between CMB 
anomalies appear. 
The magnetic f\/ield scenario analysed here is just the simplest one, and 
considering more realistic f\/ields new ef\/fects could be present on the CMB 
temperature and polarization data 
(see, e.g.,~\citealt{b95,b96,b97,b102,b100,b98,b99}, and references therein). 
In the same form, as well, possible relationships between these new ef\/fects and 
statistical\/ly anisotropic phenomena found in WMAP data could be established, all 
this deserving a better investigation.

%%%%%%%%%%%%%%%%%%%%%%%%%%%%%%%%%%%%%%%%%%%%%%%%%%%%%%%%%%%%%%%%%%%%%%%%%%%%%%%%%%
\section*{Acknowledgments}
%%%%%%%%%%%%%%%%%%%%%%%%%%%%%%%%%%%%%%%%%%%%%%%%%%%%%%%%%%%%%%%%%%%%%%%%%%%%%%%%%%
\noindent
We are grateful for the use of the Legacy Archive for Microwave Background Data 
Analysis (LAMBDA). 
We also acknowledge the use of CMBFAST (http://www.cmbfast.org) developed by 
U. Seljak and M. Zaldarriaga. 
Some of the results in this paper have been derived using the HEALPix 
package~\citep{b93}. 
WSHR acknowledges f\/inancial support from the Brazilian Agency CNPq, process 
150839/2007-3; AB acknowledges a PCI/DTI (MCT-CNPq) fellowship. 
We thank T. Villela, C.A. Wuensche, I.S. Ferreira and G.I. Gomero for 
insightful comments and suggestions.
WSHR is grateful to J.C. Fabris and the Gravitation and Cosmology Group at UFES 
for the opportunity to work there. 

%%%%%%%%%%%%%%%%%%%%%%%%%%%%%%%%%%%%%%%%%%%%%%%%%%%%%%%%%%%%%%%%%%%%%%%%%%%%%%%

%%%%%%%%%%%%%%%%%%%%%%%%%%%%%%%%%%%%%%%%%%%%%%%%%%%%%%%%%%%%%%%%%%%%%%%%%%%%%%
\label{lastpage}

\end{document}